# Tensor-Engineered Van der Waals $NbOCl_2$ Resonant Metasurface for Polarization-entangled Bell State Generation

Xin Zeng[1,2,#], Wenna Du[1,7,#], Yun-Kun Wu[3,8,#], Yuyang Zhang[4], Zhiyong Zhang[1], Kwok Kwan Tang[4], Ruihuan Duan[5], Xiaotian Bao[1], Yiyang Gong[1,7], Yuexing Xia[1,7], Yutong Zhang[1,7], Jinhan Zhao[1], Yubo Tian[1], Yangguang Zhong[1], Chuanxiu Jiang[1], Yubin Wang[1,7], Jianhui Fu[1,7], Shuai Yue[1,7], Guankui Long[2], Qing Zhang[4,*], Zheng Liu[5,6,*], Xi-Feng Ren[3,8,*], and Xinfeng Liu[1,7,9,*]

[1]CAS Laboratory of Standardization and Measurement for Nanotechnology, National Center for Nanoscience and Technology, Beijing 100190, P.R. China

[2]Frontiers Science Center for New Organic Matter, Tianjin Key Lab for Rare Earth Materials and Applications, Renewable Energy Conversion and Storage Center (RECAST), School of Materials Science and Engineering, National Institute for Advanced Materials, Nankai University, Tianjin, 300350, P.R. China

[3]CAS Key Laboratory of Quantum Information, University of Science and Technology of China, Hefei, 230026, P.R. China

[4]School of Materials Science and Engineering, Peking University, Beijing 100871, P.R. China

[5]School of Materials Science and Engineering, Nanyang Technological University, Singapore, Singapore

[6]School of Electrical and Electronic Engineering, Nanyang Technological University, Singapore, Singapore

[7]University of Chinese Academy of Sciences, Beijing, 100049, P.R. China

[8]Hefei National Laboratory, University of Science and Technology of China, Hefei, 230088, P.R. China

[9]Beijing Key Laboratory of Advanced Micro- and Nanoscale Light Source Materials and Devices, Peking University, Beijing 100871, P. R. China

[#]Xin Zeng, Wenna Du and Yun-Kun Wu contributed equally to this work.

[*]E-mail: liuxf@nanoctr.cn; renxf@ustc.edu.cn; z.liu@ntu.edu.sg; q_zhang@pku.edu.cn;

**Abstract:**

Polarization-entangled photon pairs are essential resources for quantum information technologies, yet realizing compact sources with intrinsically controllable entanglement remains challenging. Van der Waals (vdW) nonlinear materials such as $NbOCl_2$ provide atomically thin platforms for quantum light generation, yet their native crystalline anisotropies as natural materials impose limitations on accessible quantum states. Here, we develop a resonant vdW nonlinear metasurface based on $NbOCl_2$ that exploits its intrinsic optical anisotropy to engineer polarization-dependent nonlinear responses. The anisotropic optical dispersion enables selective manipulation of resonant modes, resulting in a three-order-of-magnitude enhancement of the nonlinear response along the c axis. By further tailoring these resonant modes, we redistribute the effective second-order nonlinear susceptibility tensor between orthogonal polarization channels, balancing the spontaneous parametric down-conversion pathways along the b and c axes. This enables polarization-entangled photon generation with a measured fidelity of up to 92%. Our work establishes metasurface-enabled nonlinear optical engineering as a strategy for enhancing and controlling quantum light generation in vdW materials, providing a pathway toward scalable quantum photonic platforms.

**Introduction:**

Entangled quantum light sources form a fundamental building block for quantum information acquisition, transmission, and processing, underpinning a wide range of applications including quantum communication[1–4], quantum computing[5–7], and quantum imaging[8,9]. Such sources are commonly realized through spontaneous two-photon emission[10], four-wave mixing[11], and spontaneous parametric down-conversion (SPDC)[12–14]. Among these approaches, SPDC is one of the most widely used methods for generating entangled photon pairs, supporting entanglement in polarization[12,15–22], energy-time[23], space[24,25] and momentum[26] degrees of freedom, thereby offering a scalable route to large Hilbert spaces. In particular, polarization-entangled quantum light sources are highly valued in quantum technologies due to their versatility and ease of manipulation[21,27]. Hitherto, polarization-entangled photon paris have been extensively demonstrated in conventional bulk nonlinear crystals, such as β-barium borate (BBO), potassium titanyl phosphate (KTP), lithium niobate ($LiNbO_3$) and gallium phosphide (GaP) have been widely studied[27–30]. However, these bulk nonlinear materials are inherently limited in scalability and integration capability, restricting their use in compact and adaptable on-chip quantum photonic systems. Therefore, the development of subwavelength-scale polarization-entangled quantum light sources that are more efficient, controllable, and integrable is of critical importance.

Van der Waals (vdWs) materials have recently attracted widespread attention due to their large nonlinear coefficients, subwavelength thickness, and relaxed phase-matching constraints[31–35]. Among them, hexagonally stacked vdWs offers a particular attractive platform for miniaturized and integrable photonic devices because of their atomically thin nature, exceptional nonlinear optical properties, and compatibility with on-chip photonic architectures[7,36,37]. However, in the energetically favored and most stable multilayer stacking configurations, the restoration of inversion symmetry can strongly suppress second-order nonlinear processes, thereby limiting their scalability for quantum optical applications[38]. $NbOCl_2$ has recently emerged as a promising platform for ultrathin quantum light sources, with significant potential for realizing ultra-compact on-chip quantum photonic devices, owing to its large second-order nonlinearity, low background noise, and high scalability[22,31]. However, despite these advantages, the symmetry constraints of the intrinsic second-order nonlinear susceptibility tensor in $NbOCl_2$, restrict the accessible nonlinear interaction channels, preventing direct generation of polarization-entangled photon states, and thereby limiting its application in quantum light sources[16,22,39,40]. In addition, the subwavelength thickness of vdWs materials inherently limits the effective nonlinear interaction length, resulting in reduced quantum light generation efficiency, which remains a key challenge for practical device implementation.

Metasurfaces, with their exceptional ability to tailor light-matter interactions at the nanoscale, have opened new opportunities for realizing compact quantum-entangled light sources[21,41,42]. By relaxing conventional phase-matching requirements, enabling precise manipulation of local electromagnetic fields, and strongly enhancing optical field confinement at selected frequencies, metasurfaces provide an effective platform for boosting nonlinear optical processes in subwavelength volumes[25,42–46]. Here, we develop a metasurface-enabled van der Waals quantum platform (vdW-QP) based on anisotropic $NbOCl_2$ that supports polarization-selective optical resonances, allowing deterministic control of its effective second-order nonlinear response. The engineered resonant modes dramatically enhance the nonlinear interaction strength, leading a three-orders-of-magnitude enhancement of second-harmonic generation (SHG) along the otherwise weak nonlinear axis. More importantly, by tailoring the effective second-order nonlinear susceptibility experienced by orthogonal polarization channels, we balance the two SPDC generation pathways required for polarization entanglement. As a result, the vdW-QP directly generates polarization-entangled Bell states without the need for external interferometric superposition or active tuning, achieving a state fidelity of $0.92 \pm 0.02$ and concurrence of $0.91 \pm 0.04$. Our work establishes a general strategy for overcoming symmetry-imposed constraints on nonlinear quantum light generation through metasurface engineering, providing a scalable route toward efficient subwavelength quantum light sources and highly integrated quantum photonic circuits.

**Results and discussion:**

***Design of a vdW-QP for Bell state***

$NbOCl_2$ crystallizes in the C2 space group and exhibits vdWs stacking along the out of

plane $a$-axis[31,47], while showing pronounced in-plane anisotropy along two orthogonal crystal axes, $b$-axis (defined as the horizontal, H-axis) and $c$-axis (defined as the vertical, V-axis) (Supplementary **Fig. S1a**). According to crystal symmetry analysis, the nonlinear optical response of $NbOCl_2$ is predominantly governed by the tensor elements $\chi_{HHH}^{(2)}$ and $\chi_{HVV}^{(2)} = \chi_{VHV}^{(2)}$ (see Supporting **Note 1** for details). In the SPDC process, the subscripts denote, in sequence, the polarization of the pump and that of the generated photons. When the pump polarization is aligned along the H-axis, the photon-pair generation in the SPDC process is predominantly determined by the susceptibility tensor elements $\chi_{HHH}^{(2)}$ and $\chi_{HVV}^{(2)}$. In $NbOCl_2$, the second-order nonlinear susceptibility tensor $\chi_{HHH}^{(2)}$ dominates over the other allowed tensor components, corresponding to a type-0 SPDC process[16,22,31]. The generated photon pairs are therefore predominantly emitted with H polarization, which can be expressed as $|\psi^+\rangle \xrightarrow{\chi_{HHH}^{(2)}} |H_s\rangle|H_i\rangle$ (left panel of **Fig. 1a**). Here, by exploiting the pronounced refractive-index anisotropy of $NbOCl_2$ (Supplementary **Fig. S1b**), we design a vdWs metasurface (see methods and Supplementary **Fig. S2** for details) that induces polarization-selective guided-mode resonances (GMRs). The resulting anisotropic nonlinear enhancement, particularly for the V-polarized direction, balances the photon-pair generation amplitudes of the two orthogonal polarization directions while preserving their coherent superposition. As a result, polarization-entangled photon pairs are generated and can be expressed as $|\psi^+\rangle \xrightarrow{\chi_{HHH}^{(2)},\chi_{HVV}^{(2)}} \frac{1}{\sqrt{2}}\left(|H_s\rangle|H_i\rangle + |V_s\rangle|V_i\rangle\right)$ (right panel of **Fig. 1a**).

The vdW-QP is designed as a C4-symmetric array of holes with a period of p = 450 nm and hole diameter of d = 115 nm, patterned on a subwavelength-thick (205 nm) $NbOCl_2$ flake. GMRs are supported when the in-plane wavevector, determined by the incident light and photonic crystal lattice, matches the propagation wavevector of the slab modes[25,48,49]. In the weak-coupling regime, resonance modes, which enables the most efficient coupling to free-space radiation, arise when the following condition is satisfied $\frac{2\pi}{P} \pm \sin\theta_{H,V}\frac{2\pi}{\lambda} = n_{eff(H,V)}\frac{2\pi}{\lambda}$. Here, $\theta_{H,V}$ denotes the angle of the incident plane wave along the H-axes and V-axes, and $n_{eff(H,V)}$ represents the effective refractive index along the corresponding axes. Due to the pronounced difference in effective refractive indices along the H-axis and V-axis, the corresponding resonance modes are expected to exhibit distinct spectral shifts according to the equation. **Figure 1b** shows the angle-dependent transmission spectra of the vdW-QP along the V-axis (left) and H-axis (right), calculated using the finite-difference time-domain (FDTD) method. Two transverse electric (TE) modes have emerged in both H and V directions, which are identified and denoted as $TE_1$ and $TE_2$. These TE modes along the V-axis show a pronounced red shift of ~106 meV relative to the H-axis modes, arising from the anisotropy in the effective refractive indices (Supplementary **Fig. S3**). This anisotropic spectral separation enables the engineering of nonlinear optical processes along the two orthogonal axes. Here, we focus on the wavelength range where the resonances of the two orthogonal axes spectrally overlap and interact. In this regime, the V-axis supports the $TE_1$ mode near 830 nm at 0°, whereas the H-axis exhibits the $TE_2$ mode near 805 nm at 1.5°. **Figure 1c** presents the transmission spectrum along the white dashed line indicated in **Fig. 1b**, while the inset shows the corresponding electric-field distribution at the strongest resonance wavelength. Along the V-axis, a clear electric-field amplitude profile characteristic of a GMR is observed, whereas along the H-axis, the electric-field amplitude distribution exhibits the feature of a bound state in the continuum (BIC) at the corresponding resonance. Owing to the larger effective refractive index and polarization-selective coupling[50,51], the V-polarized mode supports stronger electric-field localization and reduced radiation leakage compared to the H-polarized mode, resulting in a significantly enhanced local electric-field intensity. The pronounced field enhancement at resonance will substantially boost the efficiency of nonlinear optical processes, including SHG and SPDC[45,52]. According to the Lorentz reciprocity theorem[53,54], the relationship between the general effective second-order nonlinear susceptibility tensor $\chi_{ijk}^{(2)eff}$ for the polarization combination $(i, j, k)$ and the related field distribution in the vdW-QP can be expressed as follows:

$$\chi_{ijk}^{(2)eff} = \chi_{ijk}^{(2)} \frac{\int E_i^{2\omega} E_j^{\omega} E_k^{\omega} dV_{eff}}{E_{i(in)}^{2\omega} E_{j(in)}^{\omega} E_{k(in)}^{\omega} V_{eff}} \tag{1}$$

Here, $\chi_{ijk}^{(2)}$ is the intrinsic second-order nonlinear susceptibility tensor of $NbOCl_2$, $E_i^{2\omega}$, $E_j^{\omega}$

and $E_k^{\omega}$ are local electric field components in the vdW-QP induced by incident waves $E_{i(in)}^{2\omega}$, $E_{j(in)}^{\omega}$ and $E_{k(in)}^{\omega}$ at frequencies $\omega$ and $2\omega$, respectively. $V_{eff}$ represents the volume of the integral covering vdW-QP. The $\chi_{ijk}^{(2)eff}$ is determined by the spatial overlap of the locally enhanced electric fields at $\omega$ and $2\omega$, normalized by the incident fields. Equation (1) indicates that the vdW-QP allows for the selective engineering of different components of the nonlinear susceptibility tensor. Moreover, the local field enhancement at both $\omega$ and $2\omega$ further amplifies the nonlinear optical response. Notably, the electric-field enhancement associated with the V-axis resonance is significantly stronger than that along the H-axis (**Fig. 1c**). According eq. (1), the stronger field confinement associated with the V-polarized resonance therefore leads to a pronounced increase in the effective tensor component $\chi_{HVV}^{(2)eff}$, while the contribution from $\chi_{HHH}^{(2)eff}$ remains modest enhancement due to the weaker field enhancement along the H-axis. This demonstrates that metasurface engineering provides an effective route for selectively redistributing the nonlinear polarization tensor response.

***Enhanced nonlinear response along the V-axis of the vdW-QP***

Based on the above design, we experimentally investigate the GMR-enhanced nonlinear response along the V-axis. The vdW-QP was fabricated by focused ion beam milling, as described in the Methods. **Figure 2a** shows a scanning electron microscopy (SEM) image of the vdW-QP with lateral dimensions of approximately 23×23 μm$^2$. The inset shows a top view of the vdW-QP with a thickness of approximately 205 nm, as confirmed by the atomic force microscopy (AFM) image in Supplementary **Fig. S4**. We define the filling factor $F$ as the ratio of the hole diameter to the period ($d/p$). As detailed in Supplementary **Fig. S5** and **Note 2**, the GMR with a Fano line-shape, allows the resonance energy to be tuned via the filling factor, with optimal quality factor (Q) of 79 observed at $F$ = 0.255 (Supplementary **Fig. S6**). This tunability enables precise control over the spectral position and linewidth of the resonance. **Figure 2b** shows the angle-resolved transmission spectrum of the vdW-QP along the V-axis at $F$ = 0.255, where the GMR at 830 nm shows excellent agreement with the simulated results.

Owing to the quantum-classical correspondence, degenerate SPDC can be viewed as the inverse process of SHG[21,55], allowing us understand the former and the second-order nonlinear response from SHG. The SHG measurements are performed using a wavelength-tunable femtosecond pulsed laser (Chameleon Ultra II, Coherent) with a pulse duration of 140 fs and a repetition rate of 80 MHz in a self-built SHG reflection optical path (see methods and Supplementary **Fig. S7** for details). The V-axis nonlinear response of the vdW-QP and $NbOCl_2$ were characterized at a resonance wavelength of 830 nm, as shown in **Fig. 2c**. The SHG from the vdW-QP is enhanced by nearly three orders of magnitude relative to $NbOCl_2$, corresponding to an effective second-order nonlinear coefficient $\chi_{eff}^{(2)}$ of 3212 pm $V^{-1}$ (see Supplementary **Note 3** for details). The normalized conversion efficiency of the vdW-QP reaches $2 \times 10^{-6}$ $cm^2$/GW, which is comparable to or even exceeds that of previously reported metasurfaces (Supplementary **Table 1**). The inset in **Fig. 2c** shows the optical image of a stronger SHG in the vdW-QP under the same power density excitation. The power-law fit reveals a quadratic dependence of the SHG intensity on the pump power, which is consistent with the expected behavior of second-order nonlinear processes (Supplementary **Fig. S8a**). **Figure 2d** compares the polarization-dependent SHG responses of $NbOCl_2$ and the vdW-QP. While $NbOCl_2$ exhibits stronger SHG along the H axis due to the $\chi_{HHH}^{(2)eff}$ tensor component, the vdW-QP reverses this anisotropy and achieves maximum SHG under V-polarized excitation through the GMR-enhanced nonlinear response. The normalized polarization-resolved SHG responses for the parallel and perpendicular components of the vdW-QP and $NbOCl_2$ are further shown in Supplementary **Fig. S9**. By fitting the polarization dependence shown in the **Fig. 2d**, the ratio $\chi_{HVV}^{(2)eff} / \chi_{HHH}^{(2)eff}$ is estimated to be approximately 0.24 for $NbOCl_2$ and 25.07 for vdW-QP (see Supplementary **Note 1** for details). The increased ratio indicates that the vdW-QP significantly enhances the contribution of the $\chi_{HVV}^{(2)eff}$ tensor component, which is beneficial for the generation of V-polarized photon pairs in SPDC. The second harmonic enhancement factor is defined as the ratio of the second harmonic intensity of the vdW-QP to that of $NbOCl_2$. The dependence of the second harmonic enhancement factor on the excitation wavelength is shown in **Fig. 2e** and Supplementary **Fig. S8b**. Under excitation at the 830 nm, the enhancement factor reaches up to 2400, highlighting the outstanding nonlinear enhancement performance compared to that of previously reported two-dimensional and bulk metasurface platforms (**Fig. 2f**). By adjusting the filling factor *F*, SHG can be enhanced over a broad wavelength range, enabling effective control of the nonlinear response along the V-axis (Supplementary **Fig. S10 and S11**). These results demonstrate that the vdW-QP constitutes an effective platform for fine-tuning anisotropic nonlinear optical responses, thereby laying a solid foundation for the realization of polarization-controlled quantum light sources.

### ***Engineered anisotropic $\chi^{(2)}$ tensor of the vdW-QP***

To further validate our design strategy for realizing polarization-entangled Bell states, the GMR along V-axis is tuned to 810 nm, to match operating wavelength of our experiments. Firstly, we investigate the second-order nonlinear response over a spectral window of approximately 50 nm to elucidate the underlying evolution. **Figure 3a-e** present the polarization-resolved SHG responses of the vdW-QP at the pump wavelength range of 780-820 nm. This behavior is governed by the associated resonant modes, as shown in Supplementary **Fig. S12**, which correspond to the $TE_1$ mode along the V-axis and the $TE_2$ mode along the H-axis, respectively. When pumped at 780 nm, the polarization-resolved SHG response (**Fig. 3a**) exhibits a characteristic two-lobed pattern with maxima along 0° and 180° (H-axis), indicating that the nonlinear response is dominated by the $\chi_{HHH}^{(2)eff}$ tensor component. Compared to the intrinsic polarization response of $NbOCl_2$ (**Fig.2d**), the pattern is elongated along the H-axis, reflecting the resonance-enhanced nonlinear response in this direction. As the pump wavelength increases and approaches the V-axis resonance (**Fig. 3b-d**), the SHG polarization pattern gradually evolves into a four-lobed structure. The V-polarized SHG intensity gradually increases to a level comparable to or exceeding that of the H-axis, reflecting an enhanced contribution from the $\chi_{HVV}^{(2)eff}$ tensor component in the vdW-QP. When the pump wavelength far away the V-axis resonance (around 820 nm), the field enhancement along the V-axis diminishes, and the SHG polarization pattern approaches a more isotropic, nearly circular distribution. These results demonstrate that the vdW-QP enables effective tailoring of the second-order nonlinear response, allowing dynamic redistribution between the $\chi_{HHH}^{(2)eff}$ and $\chi_{HVV}^{(2)eff}$ contributions. A pronounced enhancement of the SHG intensity is observed along the V-axis at the corresponding resonance wavelength (Supplementary **Fig. S13**), while the SHG response along the H-axis exhibits a relatively weaker enhancement. The wavelength-dependent SHG intensity ratio $I_{HVV}/I_{HHH}$ between the two orthogonal axes in the vdW-QP and $NbOCl_2$ showed in **Fig. 3f**. For $NbOCl_2$, the intrinsic symmetry constraints of its second-order nonlinear susceptibility tensor lead to $I_{HVV}$ being much weaker than $I_{HHH}$ over the entire selected spectral range. Owing to the quantum-classical correspondence between SHG and degenerate SPDC, the pronounced SHG anisotropy in the $NbOCl_2$ corresponds to strongly imbalanced SPDC amplitudes, preventing efficient polarization-entangled state generation. By contrast, the vdW-QP redistributes the nonlinear response such that $I_{HVV}$ becomes comparable to, or even exceeds, $I_{HHH}$ over a broad spectral range, thereby enabling balanced SPDC pathways required for Bell-state generation.

To predict the performance of polarization-entangled photon pairs, the fidelity $F(\lambda_0)$ and the concurrence $C(\lambda_0)$ of the generated entangled state can be calculated by $F(\lambda_0) = \left(1 + R(\lambda_0)\right)^2/2[1 + R(\lambda_0)^2]$ and $C(\lambda_0) = 2R(\lambda_0)/[1 + R(\lambda_0)^2]$ (see Supplementary **Note 4** for details). Here, $R(\lambda_0)$ is defined as the ratio of the amplitudes along the H- and V-axes, each of which is governed by the local electric field intensity and the corresponding second-order nonlinear susceptibility tensor component. To simplify the calculation, we consider a degenerate SPDC process with an idealized zero bandwidth, allowing the entanglement properties to be evaluated independently at each wavelength. We calculate the integrated electric field intensity of H-axis and V-axis within the selected spectral range (Supplementary **Fig. S14**). As the SPDC wavelength approaches the V-axis resonance region, the GMR strongly enhances the local electric field. Using above equations, we evaluate the Bell-state fidelity as a function of the central wavelength, as shown in **Fig. 3g**. It presents the simulated wavelength-dependent fidelity $F(\lambda_0)$ and concurrence $C(\lambda_0)$ of the generated polarization-entangled state in the vdW-QP. Both quantities exhibit pronounced wavelength-dependent oscillatory behavior and reach their highest values on either side of the V-axis resonance peak (810 nm). At these points, they approach unity, indicating the formation of nearly maximally entangled Bell states. As the wavelength is detuned from the V-axis resonance towards the H-axis resonance, the fidelity and concurrence decrease markedly, reflecting a degradation of the entanglement quality. The simplified calculation provides an effective method for predicting the feasibility of polarization-entangled photon generation. These results further demonstrate that engineering the V-axis GMR is crucial for achieving polarization-entangled Bell-state generation.

***Polarization-entangled Bell state generation in the vdW-QP***

We characterized the quantum emission properties of the vdW-QP using a home-built Hanbury Brown-Twiss (HBT) setup, as schematically shown in **Fig. 4a**. A 405 nm continuous-wave laser was used as the pump source, and the emitted signals were collected through two independent detection channels. Each channel was equipped with a quarter-wave plate, a half-wave plate, and a linear polarizer, enabling projective measurements onto arbitrary polarization states.

The SPDC emission along the V-polarization direction in the vdW-QP was systematically investigated. **Figure 4b** shows the normalized second-order correlation function $g^{(2)}(\tau)$ as a function of pump power. The inset displays the zero-delay second-order correlation $g^{(2)}(0)$ as a function of pump power. Notably, $g^{(2)}(0)$ remains well above 2 over the entire pump-power range and increases with decreasing pump power. This inverse dependence is the characteristic of spontaneous photon-pair generation, as the reduced pump power suppresses multi-pair emission and accidental coincidences, thereby enhancing the measured photon correlations[31]. These results clearly demonstrate enhanced photon-pair emission from the V-polarized SPDC channel in the vdW-QP. The $g^{(2)}(\tau)$ measurements of the H-polarized SPDC channel under different pump powers are shown in Supplementary **Fig. S15a-b**.

**Figure 4c** shows the $g^{(2)}(\tau)$ signals measured at a pump power of 1 mW for the vdW-QP and $NbOCl_2$ along both orthogonal axes (integration time 30 minutes). In the vdW-QP, a pronounced enhancement of SPDC emission is observed along the V-axis, while only a weak enhancement occurs along the H-axis. In contrast, the SPDC along the V-axis in $NbOCl_2$ is negligible. The corresponding coincidence counts (**Fig. 4d**) exhibit a linearly dependence on the pump power, confirming the spontaneous nature of the photon-pair generation process. Notably, comparable coincidence rates are obtained in the HH and VV bases for the vdW-QP, satisfying the amplitude-balance condition required for the generation of polarization-entangled Bell states.

To directly verify the generation of polarization-entangled Bell states, we performed quantum state tomography to reconstruct the full density matrix of the generated photon pairs[16,22,56]. The real and imaginary parts of the reconstructed density matrix are shown in **Fig. 4e**. In the 4×4 density matrix, significant populations are observed only in the HH and VV bases, while the orthogonal bases are strongly suppressed, consistent with an ideal Bell state $|\psi\rangle$ (**Fig. 4f**). By comparison with the ideal Bell state density matrix, the reconstructed state from vdW-QP exhibits a fidelity of 0.92 ± 0.02, in close agreement with the theoretical expectation. In addition, a concurrence value of 0.91 ± 0.04 is obtained, confirming the generation of high-quality polarization-entangled photon pairs.

**Discussions:**

In conclusion, we have demonstrated the direct generation of polarization-entangled Bell states in a metasurface-enabled vdW-QP through effective nonlinear response engineering. By tailoring the filling factor of the vdW-QP, broadband control over the polarization-dependent second-order nonlinear response is achieved, resulting in a maximum nonlinear enhancement approaching three orders of magnitude. The GMR selectively amplifies the originally weaker nonlinear polarization channel, thereby rebalancing the anisotropic nonlinear interaction and enabling polarization-entangled photon-pair generation in a system whose intrinsic nonlinear susceptibility would otherwise preclude direct polarization entanglement. Consequently, an ultracompact polarization-entangled quantum light source is realized, exhibiting a Bell-state fidelity of 0.92 ± 0.02 and a concurrence of 0.91 ± 0.04.

Beyond advancing the application of $NbOCl_2$ in integrated quantum photonics, this work establishes a general strategy for tailoring the effective nonlinear response of anisotropic vdWs materials through metasurface engineering. By combining the intrinsic scalability and atomically thin nature of layered materials with the versatile control of light-matter interaction enabled by dielectric metasurfaces, our platform provides a compact, efficient and highly flexible route toward integrated entangled-photon sources without requiring external interferometric architectures, stringent phase-matching conditions, or precise interlayer stacking engineering.

Looking forward, the resonance characteristics of the vdW-QP can be continuously tuned through geometric engineering, enabling independent control of multiple resonant modes across different wavelengths and polarization channels within a single device. Such flexibility offers new opportunities for simultaneously manipulating multiple optical degrees of freedom and generating more complex quantum photonic states. In particular, when combined with the intrinsic polarization anisotropy of $NbOCl_2$, this platform provides a promising pathway toward polarization-frequency hyperentanglement and multifunctional on-chip quantum photonic devices.

## Methods Section

**Materials and fabrications.** Bulk single-crystalline $NbOCl_2$ were fabricated by chemical vapor transport[31]. Next, $SiO_2$/Si substrates were meticulously cleaned in an ultrasonic bath filled successively with ethanol, acetone, and deionized water for 15 min each. Thin $NbOCl_2$ flakes were prepared on a clean $SiO_2$/Si substrate via the mechanical exfoliation method using adhesive tape. AFM measurements were performed to determine the thickness of the samples.

The periodic air-hole lattices were fabricated by FIB milling. Using the Helios 5 UX system (Thermo Fisher Scientific), $23 \times 23$ $\mu m^2$ arrays of periodic air holes were patterned on the $NbOCl_2$ flake. To obtain an appropriate etching spot size, the $Ga^+$ ion beam current was maintained below 41 pA. Etching paths of ion beam scanning on the $NbOCl_2$ microplatelet was optimized using the stream file function in the FEI system, and the fabrication dwell time was set as a maximum of 80 ms. The processing procedure is shown in Supplementary **Fig. S2**. SEM images were acquired using the same system, capturing views at 0° and 52° inclinations with an electron beam (5 KV). Finally, a layer of PMMA was spin-coated onto the processed sample at 3000 rpm for 30 s to provide refractive index matching and surface protection.

**Angle-Resolved Spectroscopy.** To obtain angle-resolved transmission spectra, a custom-made Fourier imaging system was utilized, incorporating an objective lens (50×, numerical aperture of 0.8) to focus light and improve spatial resolution. In reflection measurement, the white light beam from a tungsten-halogen light source (SLS201L, Thorlabs) was focused with the objective lens onto the sample. The reflected light was collected by a Princeton Instruments spectrometer (SP-2500i) equipped with a liquid-nitrogen-cooled charge-coupled device detector and gratings of 150 and 600 grooves per mm.

**SHG Characterization.** SHG measurements were carried out in a reflection geometry through our home-built micro-area nonlinear optical characterization system (Supplementary **Fig. S7a**). Pump pulses from a Ti:sapphire oscillator (Coherent, Chameleon Ultra II; 800 nm central wavelength, tunable from 680 to 1080 nm; 140 fs pulse duration; 80 MHz repetition rate) were focused onto the vdW-QP using an Olympus objective lens (MPlanFL N, 20×/0.45 BD), resulting in a spot size of ~12.5 μm at the pump wavelength. The SHG signal was then back collected by the same lens, separated using a dichroic mirror, and filtered by a short pass filter before entering the Princeton Instruments spectrometer (SP-2500i). Polarization measurement was carried out by the combination of a half-waveplate (HWP, linear polarization) with a linear polarizer before the sample.

**SPDC and quantum tomography.** All SPDC measurements were performed using a home-built Hanbury Brown-Twiss setup, as schematically shown in **Fig. 4a**. A broadband white-light source (SLS201L, Thorlabs) was focused onto the vdW-QP through an objective lens for imaging and alignment. All measurements were carried out in a reflection configuration.

A continuous-wave 405 nm laser was used as the pump source and was focused onto the vdW-QP by an objective lens (MPlanFL N, 20×/0.45 BD). The generated SPDC signal was collected by the same objective. A dichroic mirror was employed to reflect the pump beam while transmitting the down-converted photons. An additional long-pass filter (LP, 750 nm) was inserted to further suppress any residual pump leakage. The reflected SPDC signal was subsequently split by a 50:50 beam splitter and coupled into two multimode optical fibers (50 μm core diameter, LBTEK). Each output channel was filtered by a band-pass filter centered at 800 nm with a bandwidth of ±25 nm before being detected by two single-photon counting modules (SPCM-AQRH-16-FC, Excelitas Technologies). Coincidence measurements and data acquisition were performed using a time-correlated single-photon counting system.

The polarization-entangled Bell state was characterized using quantum state tomography[16,22]. Based on the optical configuration described above, a complete set of 16 polarization projection measurements was performed using combinations of a quarter-wave plate, a half-wave plate, and a linear polarizer to reconstruct the two-photon quantum state. The corresponding density matrix was then reconstructed from the measured data using a maximum-likelihood estimation method.

**Numerical simulation methods.** We used the commercial software Lumerical FDTD to perform the mode dispersions and electric field amplitude distribution. All the geometric parameters in the simulations were taken from the SEM images and AFM images. The wavelength-dependent complex refractive index of $NbOCl_2$ was obtained from the literature[47].

**Data availability**

All data to evaluate the conclusions are present in the manuscript, and the Supplementary Material. Raw data are available from the corresponding authors on request.

**Associated content**

*Supplementary Information

The Supplementary Information is available.

Supplementary Figs. S1-S15 and Notes 1-4.

**Author information**

Corresponding Authors

*Email: liuxf@nanoctr.cn; renxf@ustc.edu.cn; z.liu@ntu.edu.sg; q_zhang@pku.edu.cn;

**Author contributions**

X.L. conceived the idea. X.L., X.R., Q.Z. lead the project. X.Z., W.D. and Y.W. perform the optical spectroscopy and data analysis. X.Z., Y.Z. and Z.Z. perform numerical simulations. X.Z., Y.W., Y.G., S.Y., J.F., C.J. and Y.Z. construct the optical set up. X.Z., Y.X., G.D. and Y.W. perform the SPDC measurement. X.Z., Y.W., Y.T., Y.Z. and Y.Z. perform the SHG measurement. X.Z., R.D., Z.Z., J.Z. and X.B. produced the samples. X.Z., W.D. and X.L. prepared the manuscript. All the authors discussed the results and revised the manuscript.

**Acknowledgments**

The authors acknowledge the support from the National Key Research and Development Program (2023YFA1507002, 2024YFA1208203, 2023YFA1407004, 2022YFA1204300 and 2022YFA1204704), National Science Foundation for Distinguished Young Scholars of China (22325301, T2325022), the National Natural Science Foundation of China (52221001, 52402181, U23A2074, 22573023). The National Science Fund for Excellent Young Scholars of China (62422502). The Natural Science Foundation of Hunan Province (2022JJ30167). Project of Yuelushan Center for Industrial Innovation (521025047). the CAS Project for Young Scientists in Basic Research (No. YSBR-049), Quantum Science and Technology-National Science and Technology Major Project (Grants 2021ZD0303200, 2021ZD0301500), and the Fundamental Research Funds for the Central Universities. S.Y. appreciates Youth Innovation Promotion Association CAS and CAS Project for Young Scientists in Basic Research, under grant no.

YSBR-120. This work was partially carried out at the USTC Center for Micro and Nanoscale Research and Fabrication.

**Competing interests:** Authors declare no competing interests.

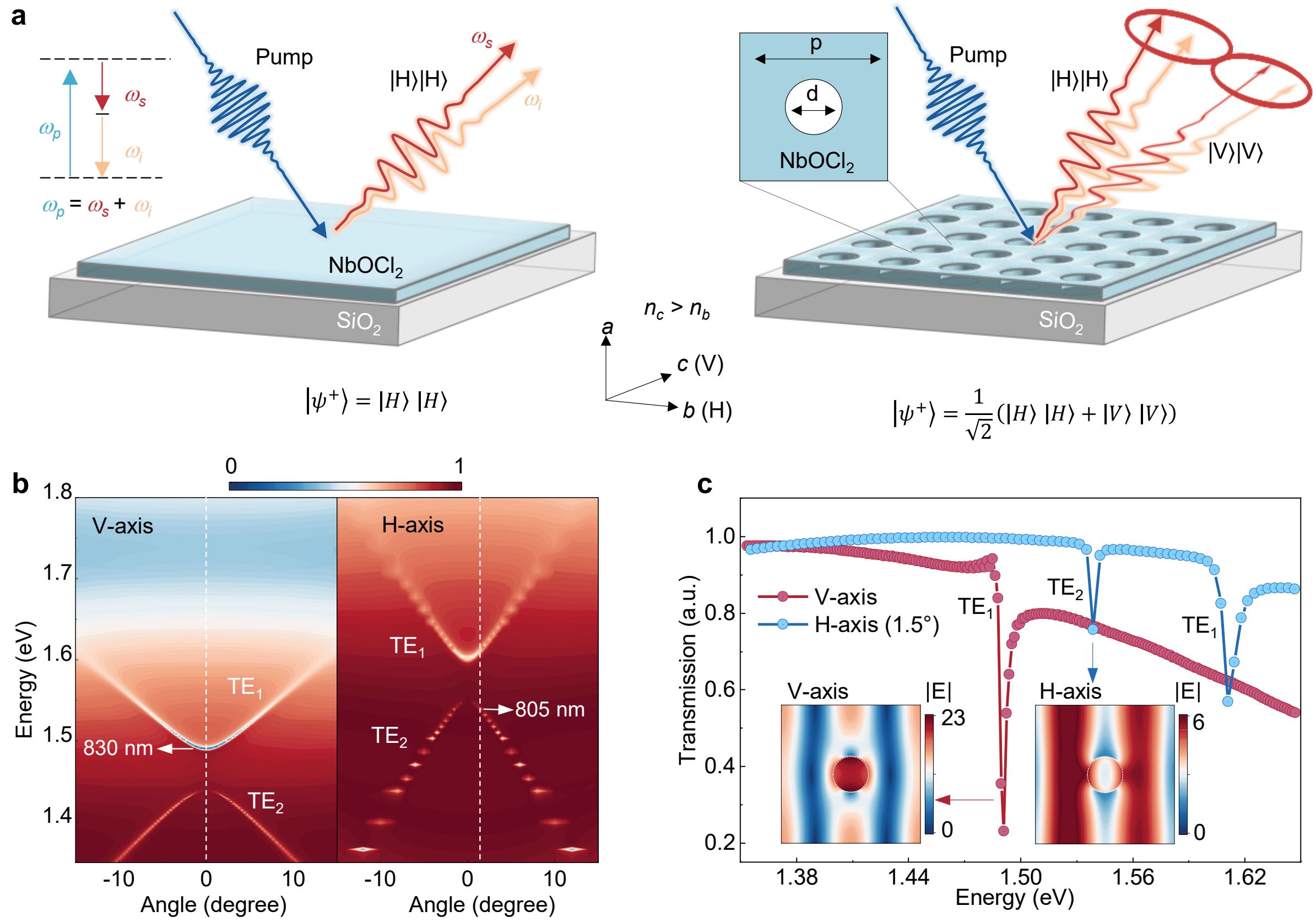

**Fig.1| Design of vdW-QP for polarization-entangled quantum light generation. a,** Conceptual illustration of down-conversion emission in $NbOCl_2$ (left) and polarization-entangled down-conversion light generation enabled by the vdW-QP (right). $NbOCl_2$ exhibits strong in-plane anisotropy, with the refractive index along the c-axis ($n_c$) larger than that along the b-axis ($n_b$). For clarity, the b-axis and c-axis are denoted as the H-axis and V-axis, respectively. Under pumping at frequency $\omega_p$, SPDC generates signal $\omega_s$ and idler $\omega_i$ photon pairs satisfying $\omega_p = \omega_s + \omega_i$. For $NbOCl_2$, both photons are H-polarized. Through metasurface-induced resonances, the SPDC emission amplitudes from the two orthogonal polarization directions can be balanced, enabling the generation of photon pairs in the Bell state$1/\sqrt{2}\,(|H\rangle|H\rangle + |V\rangle|V\rangle)$. **b,** Calculated angular resolved transmission spectra of vdW-QP along the V-axis (left) and H-axis (right) under TE polarization. The maximum electric field enhancement is observed along the V-axis at 830 nm ($TE_1$), in contrast to 805 nm ($TE_2$) for the H-axis. **c,** Transmission spectra of the vdW-QP calculated under plane-wave excitation, polarized along the V-axis at normal incidence (red) and along the H-axis at an incident angle of 1.5° (blue). The inset shows the electric field amplitude distributions along the V-axes and H-axes at the corresponding resonant wavelengths.

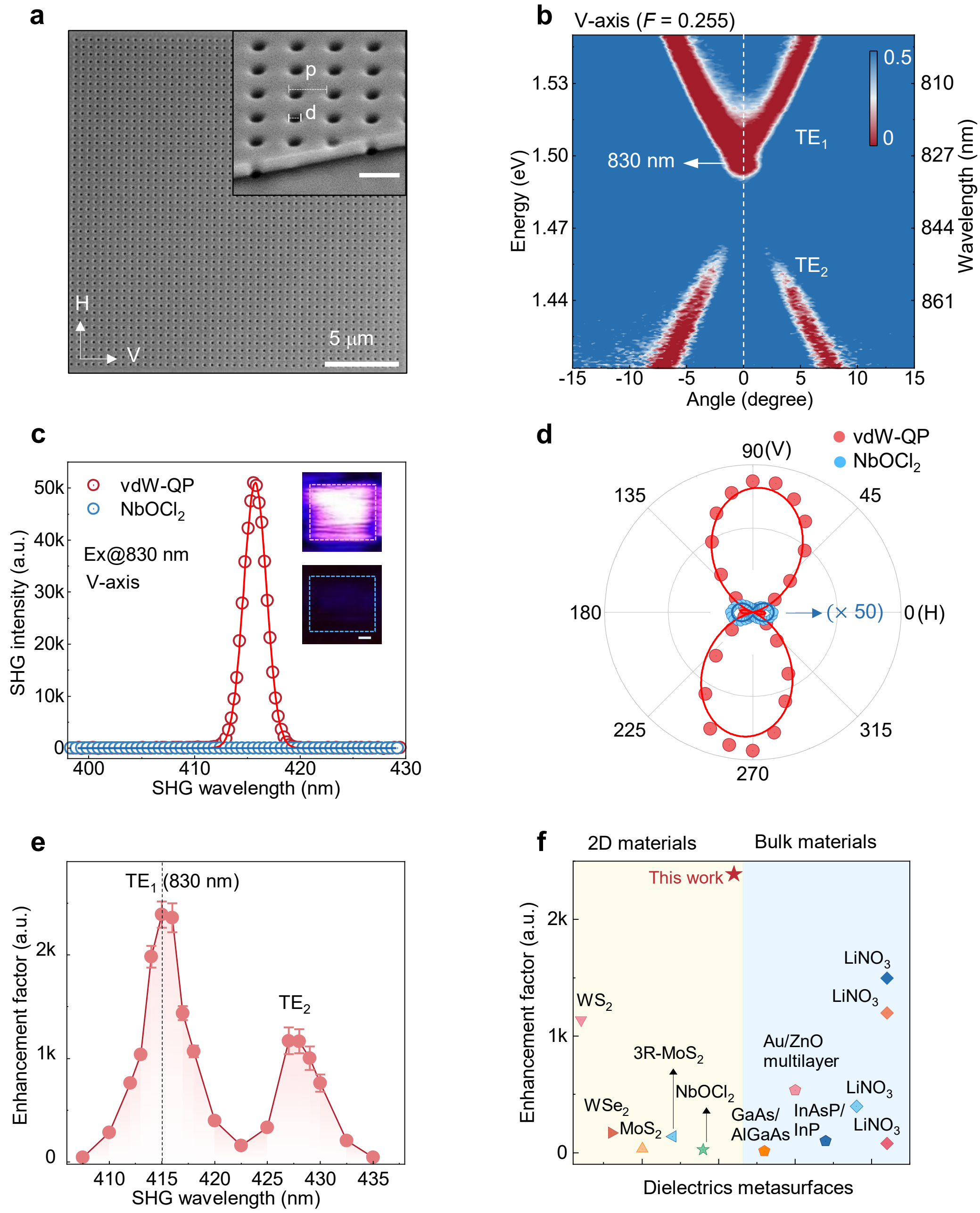

**Fig.2| vdW-QP for efficient SHG. a,** The top view SEM image of vdW-QP. The entire array occupies an area of approximately 23×23 μm$^2$. The illustration shows the SEM image of vdW-QP tilted at 52° (The scale bar is 500 nm). **b,** Angle-resolved TE-polarized transmission spectra of the vdW-QP along the V-axis at a filling factor $F$ = 0.255. It shows $TE_1$ and $TE_2$ modes, consistent with the simulations, and indicates the strongest SHG enhancement at 830 nm. **c,** SHG spectra collected at the same power in vdW-QP (patterned region, red line) and $NbOCl_2$ (unpatterned region, blue) at an excitation wavelength of 830 nm. Illustration: Optical images of SH signals in patterned (up) and unpatterned (down) regions at the same excitation power, the scale bar is 4 mm. **d,** Polarization-dependent SHG of vdW-QP (red) and $NbOCl_2$ (blue). **e,** Wavelength-dependent SHG enhancement factors, defined as the SH intensity from the vdW-QP normalized to that of the unpatterned flake along the V-axis. **f,** SHG enhancement factor of vdW-QP compared with other metasurfaces.

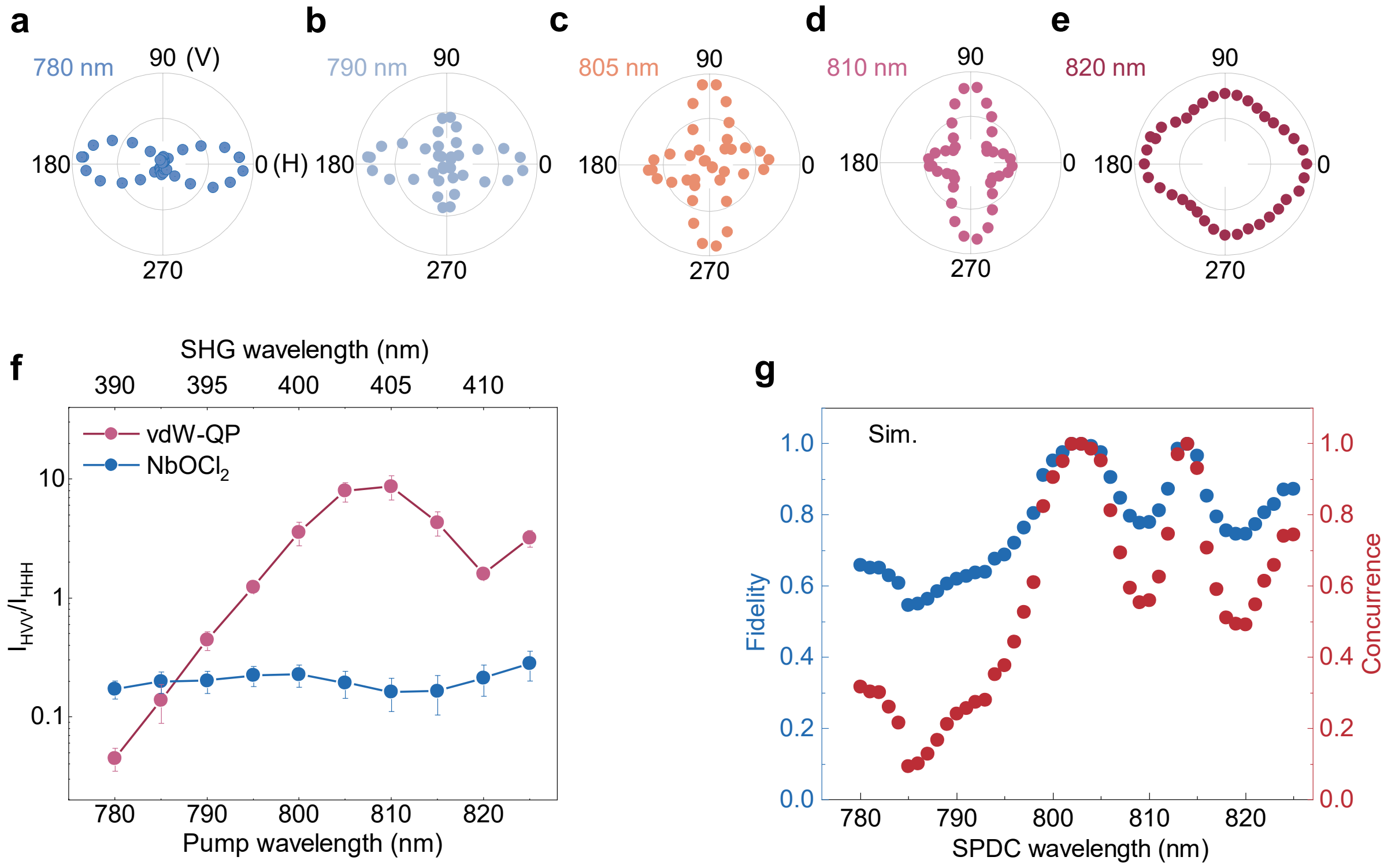

**Fig.3| Engineered anisotropic $\chi^{(2)}$ tensor of the vdW-QP. a-e,** Polar plots of the SHG intensity as a function of polarization angle measured at SHG wavelengths of 390, 395, 402, 405, and 410 nm, respectively. **f,** Wavelength-dependent SHG intensity ratio between the $I_{HVV}$ and $I_{HHH}$ processes in vdW-QP and $NbOCl_2$. As the pump wavelength approaches the V-axis resonance (810 nm) of the vdW-QP, the SHG intensity of $I_{HVV}$ can be tuned to approaches and even exceed that of $I_{HHH}$. By contrast, it shows small ratios for $NbOCl_2$ throughout the entire wavelength range. **g,** The calculated wavelength-dependent fidelity and concurrence of the Bell states in vdW-QP.

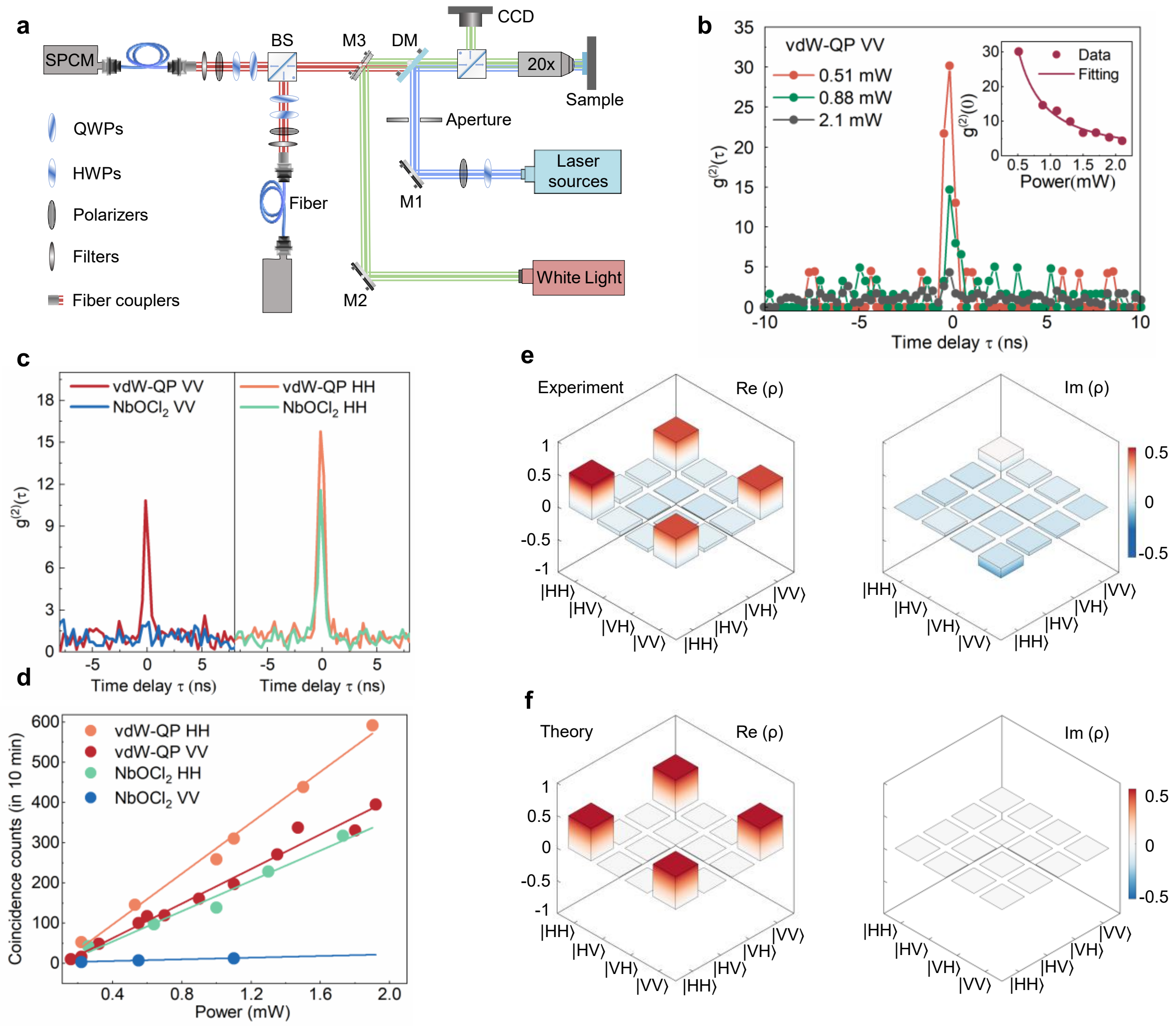

**Fig.4| Experimental demonstration and Bell-state construction with vdW-QP. a**, Schematic of the optical setup for SPDC characterization and polarization-state tomography. QWP: quarter-wave plate, HWP: half-wave plate, BS: beam splitter, DM: dichroic mirror, M: mirror, SPCM: single-photon counting module. **b,** Power dependent normalized second correlation along the V-axis polarization in the vdW-QP. The inset shows the power dependent normalized correlation function measured at zero delay. **c,** Second correlation function measured from vdW-QP and $NbOCl_2$. Detection directions are defined relative to the $NbOCl_2$ axes, with the pump polarization aligned along the H-axis to maximize the coincidence counts. The left and right panels show the SPDC signals along the V-axes and H-axes, respectively. **d,** Coincidence counts as a function of pump power at vdW-QP and $NbOCl_2$ with different polarization directions. Experimental **(e)** and theoretical **(f)** polarization density matrices $\rho$ of the vdW-QP. Left panels: real parts, right panels: imaginary parts.